\documentstyle[12pt]{article}
\begin{document}
\pagestyle{headings}
{\bf \ \ } \\[1cm]
\begin{flushright} {\bf ISU-IAP.Th94-01,\ Irkutsk} \end{flushright}
\vspace{2cm}
\begin{center}
{\Large \bf First estimates of the $(\alpha + \beta)^{\pi}$
from two--photon experiments.}\\[1cm]
{\Large A.E.Kaloshin, V.M.Persikov and
V.V.Serebryakov \footnote{Institute for Mathematics,
630090, Novosibirsk, Russia.}}
\end{center}
\begin{center}
\sl Institute of Applied Physics, \\
Irkutsk State University,\\
664003, Irkutsk, Russia\\[2cm]
\end{center}
{\large Abstract.} \\

We carry out the semi--model analysis of existing data on the angular
distributions of the
$\gamma\gamma\rightarrow\pi^+ \pi^- ,\ \pi^0 \pi^0$
reactions with purpose to obtain
S-wave cross section and D-wave's parameters. In the first time
we obtain from experiment the sum of electrical and magnetic
pion polarizabilities:
$(\alpha + \beta)^{\pi^+} = 0.28 \pm 0.07$ (MARK-II \cite{M2}), \ \
$(\alpha + \beta)^{\pi^+} = 0.38 \pm 0.05$ (CELLO \cite{CELLO}), \ \
$(\alpha + \beta)^{\pi^0} = 1.26 \pm 0.06$ (Crystal Ball \cite{CB})
\ \ in units of $10^{-42} \ cm^3$ ($e^2=4 \pi \alpha$). \\
\newpage
1. \ \ Nowadays there exist an experimental data on the angular
distributions of the $\gamma\gamma\rightarrow\pi^+ \pi^-,\
\pi^0 \pi^0$ reactions \cite{M2,CELLO,CB}. Probably the most interesting
physical question here is related with the S-wave cross section
(long--standing problem of scalar resonances spectrum and structure).
However the D--wave dominates in these reactions in a wide energy
region, so this task needs either a bid statistics or the accurate
modelling of the main contribution. Note that the D-wave
contains the low--energy parameter, not investigated earlier ---
the sum of electrical and magnetic pion polatizabilities.
The most surprising result of our analysis is the fact that
these low--energy parameters may be obtained from existing data with
small statistical errors and with minimal model assumptions.

An angular distributions can be analyzed in different ways:
from model--independent partial--wave analysis
(see \cite{Harjes}) up to using of model final state formulae
for both helicity amplitudes \cite{MP}.
Here we use some intermediate method of analysis \cite{CELLO}:
for the main helicity 2 amplitude some model is used with few
free parameters,
as for S--wave contribution -- it is extracted from data
independently in every energy bin. The reasons for such a choice
are that model--independent analysis gives too ambiguous
results \cite{Harjes} with present data and just the S-wave
contains the biggest theoretical uncertainty.
In contrast to \cite{CELLO} we  use the unitary
model \cite{KS-86} with final state interaction for helicity 2
amplitude instead of rather rough simple expression.
Its  using leads \cite{KP} to other conclusions as compared with
\cite{CELLO}: a) \ There is no necessity for additional damping
of Born QED contribution (of unclear nature) to describe the data ,
b)\ The S-wave below 1 Gev is much less and does not
conflict with results of near--threshold analysis \cite{KS-92}.
Present more detailed analysis confirms the main conclusions
of \cite{KP} and gives the additional arguments in favour of
the model \cite{KS-86}.

In such an analysis the main assumption is the dominantness
on helicity 2 state in the decay $f_2(1270) \rightarrow \gamma
\gamma$. It has theoretical foundations and does not
contradict to previous numerous experiments. The quality
of present data does not allow to check this assumption
on our analysis.\\
2.\ \ Formulae for helicity 2 amplitude are contained in
\cite{KS-86}, so we shall specify here only the background
contribution, interfering with resonance $f_2(1270)$.
\begin{eqnarray}
W^C &=& M_{+-}^{Born} +
\frac{g^2_{\rho \pi \gamma}}{4} (\frac{1}{m^2_{\rho} - t} +
\frac{1}{m^2_{\rho} - u} ) + a^C   \nonumber  \\
W^N &=& \frac{g^2_{\rho \pi \gamma}}{4} (\frac{1}{m^2_{\rho} - t} +
\frac{1}{m^2_{\rho} - u} ) + (\rho \rightarrow \omega) + a^N
\label{W}
\end{eqnarray}
Here $a^{C,N}$ are some arbitrary constants combining all other
possible contributions , C = Charged,\ N = Neutral .
The same idea was used in \cite{KS-92} for helicity 0 amplitude:
only the lightest $\rho$- and  $\omega$- cross--exchanges
are considered as "alive" , all other are approximated
by an arbitrary constant. It gives the more general formulae
with additional degrees of freedom. Note that
these amplitudes satisfy the one--channel unitary condition,
I = 0, J = 2 final state interaction we describe in a
standard way, I = 2, J = 2 interaction  we don't
take into account, as usual, because of its smallness.
All contributions near the Compton--effect threshold
contribute to the low--energy structure constants
$(\alpha + \beta)/ 2 m_{\pi}$. We prefer to use them as
free parameters instead of $a^{C,N}$ . As a result
helicity 2 amplitudes
$\gamma\gamma\rightarrow\pi^+ \pi^- ,\ \pi^0 \pi^0$
will contain three arbitrary constants:
$(\alpha + \beta)^C$, $(\alpha + \beta)^N$, $\Gamma(f_2 \rightarrow
\gamma \gamma)$. \\
3.\ \ In analysis we found that the data on
$\gamma \gamma \rightarrow \pi^+ \pi^-$
are sensitive only to polarizability of $\pi^+$ and
$\gamma \gamma \rightarrow \pi^0 \pi^0$
 only to polarizability of $\pi^0$. So in considering of a single reaction
we shall fix this unessential parameters according to
theoretical prediction and use the two--parameter expression
for helicity 2 amplitude. The fixed parameter may lie in a very
broad interval -- see below.
Let's recall that predictions of
different low--energy models (i.e. \cite{Din,VO} )
are rather close to each other:
$(\alpha + \beta)^{\pi^+} \simeq  0.20 $,
$(\alpha + \beta)^{\pi^0}  \simeq  1.20 $ in units of $10^{-42} \ cm^3$
( $e^2 = 4 \pi \alpha $. Slightly the higher values are predicted in
the dispersion sum rules \cite{Petr}:
$ 0.43 \pm 0.06 $ and \
$ 1.65 \pm 0.13 $ correspondingly.
\\
\underline {
CELLO data on $\gamma \gamma \rightarrow \pi^+ \pi^-$ \cite{CELLO}}.
Let's consider the CELLO angular distributions
in 7 energy bins from 0.8 up to 1.4 Gev, they contain 53 points.
Let us fix $(\alpha + \beta)^{\pi^0} =  1.20 $,
one can vary it between 0.6 and 1.8 without any
changes. Best fit values are:
\begin{eqnarray}
(\alpha + \beta)^{\pi^+} = 0.38 \pm 0.05,\ \ \ \ \ \ \ \ \ \ \
\Gamma(f_2(1270)\rightarrow \gamma \gamma) = 2.88 \pm 0.12 \ Kev
\nonumber \\
\chi^2 = 54 \ \ \ at  \ \ \ \ NDF = 53 - 9 = 44\ \ \ \ \ \ \ \ \ \
\ \ \ \ \ \
\end{eqnarray}
\underline {MARK-II data on  $\gamma \gamma \rightarrow \pi^+ \pi^-$
\cite{M2}}.
Seven energy intervals between 0.8 and 1.4 Gev,  42 experimental
points. Again the $(\alpha + \beta)^{\pi^0} =  1.20 $  is fixed.
\begin{eqnarray}
(\alpha + \beta)^{\pi^+} = 0.28 \pm 0.07,\ \ \ \ \ \ \ \ \ \ \
\Gamma(f_2(1270)\rightarrow \gamma \gamma) = 2.73 \pm 0.19 \ Kev
\nonumber \\
\chi^2 = 17.0 \ \ \ at  \ \ \ \ NDF = 42 - 9 = 31\ \ \ \ \ \ \ \ \ \
\ \ \ \ \ \
\end{eqnarray}
\underline {
Crystal Ball  data on $\gamma \gamma \rightarrow \pi^0 \pi^0$ \cite{CB}}.
Consider seven energy intervals between 0.85 and 1.45 Gev,
56 points. \ $(\alpha + \beta)^{\pi^+} =  0.20 $  is fixed,
its changing between \ - 0.5 and 1.0 does not influence on results.
\begin{eqnarray}
(\alpha + \beta)^{\pi^0} = 1.26 \pm 0.06, \ \ \ \ \ \ \ \ \ \
\Gamma(f_2(1270)\rightarrow \gamma \gamma) = 3.56 \pm 0.22 \ Kev
\nonumber \\
\chi^2 = 38 \ \ \ at  \  \ \ \ \  NDF = 56 - 9 = 47\ \ \ \ \ \ \ \ \ \
\ \ \ \ \ \
\end{eqnarray}
4.\ \ The main our observation in such an analysis is that
the angular distributions for both reactions in vicinity of
the resonance $f_2(1270)$ are very sensitive to value of smooth
background, interfering with resonance. Simplest assumption
about form of this background allows to relate it with a
threshold structure constant --- $(\alpha + \beta)^{\pi}$ and
to obtain in the first time some estimates for them. The
obtained values are very close to existing theoretical
predictions. At first sight there should exist the essential
model dependence at an extraction of threshold parameter
from analysis in region of $f_2(1270)$. But we found that
the  $(\alpha + \beta)^{\pi}$ values are very stable at any
attempts "to improve"  the model. In particular, one can
include the higher mass exchanges to background contribution
(\ref{W}) --- it redefines the constants $a^{C,N}$,
but does not changes polarizability's estimates.

Another interesting result is the smallness
of the S--wave cross section, which is seen in
analysis of all existing data \cite{M2,CELLO,CB}.
Only in vicinity of 1.3 Gev there arises some resonance--like
structure of a rather small amplitude. The results for three
considered experiments does not contradict to each other
in the first approximation.
Such a behaviour of the
S-wave must give additional constrains for resonance
interpretation of data in scalar sector and it needs a more
detailed consideration.

This work was supported in part by grant 2-63-8-27
of Russian State Committee on High Education.

\end{document}